\documentclass[submission, Phys]{SciPost}

\begin{document}

\begin{center}{\Large \textbf{
Exclusive electro-weak production of a charmed meson at high energy
}}\end{center}

\begin{center}
B.~Pire\textsuperscript{1},
L.~Szymanowski\textsuperscript{2},
J. Wagner\textsuperscript{2*}
\end{center}

\begin{center}
{\bf 1} CPHT, CNRS, \'Ecole Polytechnique,
 I.P. Paris, 91128 Palaiseau,     France
\\
{\bf 2} National Centre for Nuclear Research (NCBJ), Pasteura 7, 02-093 Warsaw, Poland
\\
* Jakub.Wagner@ncbj.gov.pl
\end{center}

\begin{center}
\today
\end{center}


\definecolor{palegray}{gray}{0.95}
\begin{center}
\colorbox{palegray}{
  \begin{tabular}{rr}
  \begin{minipage}{0.1\textwidth}
    \includegraphics[width=22mm]{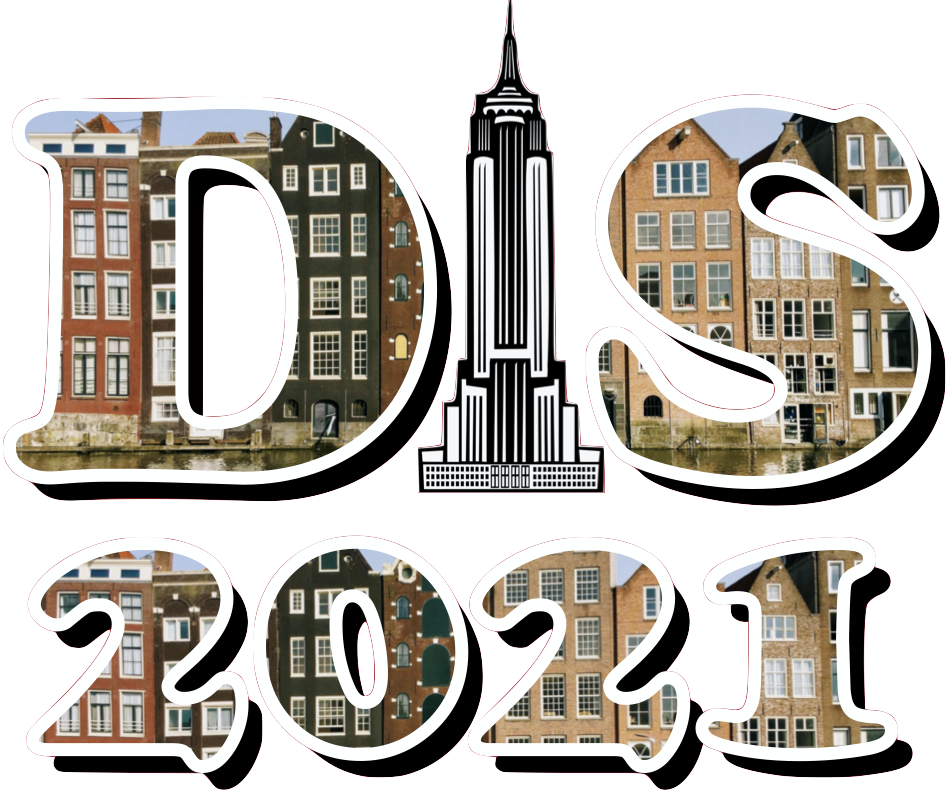}
  \end{minipage}
  &
  \begin{minipage}{0.75\textwidth}
    \begin{center}
    {\it Proceedings for the XXVIII International Workshop\\ on Deep-Inelastic Scattering and
Related Subjects,}\\
    {\it Stony Brook University, New York, USA, 12-16 April 2021} \\
    \doi{10.21468/SciPostPhysProc.?}\\
    \end{center}
  \end{minipage}
\end{tabular}
}
\end{center}

\section*{Abstract}
{\bf
Exclusive electro-weak processes have peculiar features which make them complementary to usually discussed deeply virtual electroproduction processes such as deep virtual Compton scattering or meson production (and the corresponding  crossed reactions). They allow in particular single charmed meson production, which we study in two different contexts : electroproduction at an electron ion collider and neutrino-production at a medium energy neutrino facility. We rely on the QCD collinear factorization framework where generalized parton distributions allow physicists to perform a nucleon tomography.

}

\vspace{10pt}
\noindent\rule{\textwidth}{1pt}
\tableofcontents\thispagestyle{fancy}
\noindent\rule{\textwidth}{1pt}
\vspace{10pt}

\section{Introduction}
\label{sec:intro}
 The very high luminosity anticipated at planned high energy electron ion colliders will open the physics domain of charged current events in  exclusive electroproduction processes to a detailed investigation of various interesting channels. Inspired by the pioneering works of \cite{Siddikov:2017nku,Siddikov:2019ahb} we calculate the leading order amplitude for  the following reactions :
\begin{eqnarray}
e^- (k)+N(p_1) &\to& \nu_e (k')+D_s^- (p_D)+N'(p_2) \,,\\
e^- (k)+N(p_1) &\to& \nu_e (k')+D_s^{*-} (p_D, \epsilon_D)+N'(p_2)\,,
\end{eqnarray}
on a nucleon N (proton or  neutron) target, and demonstrate that these processes may be accessed at future electron-ion colliders \cite{Accardi:2012qut,Anderle:2021wcy,AbdulKhalek:2021gbh}.

This study is much related to the reactions one may access in neutrino experiments, namely
\begin{eqnarray}
\nu_e (k)+N(p_1) &\to& e^-(k')+D^+ (p_D)+N'(p_2) \,,\\
\nu_e  (k)+N(p_1) &\to& e^-(k')+D^{*+} (p_D, \epsilon_D)+N'(p_2)\,,
\end{eqnarray}
which we discussed some time ago \cite{Pire:2017lfj, Pire:2017yge} but have not yet been confronted to experimental data.

\section{Neutrino-production of a heavy meson}
\begin{figure}[h]
\includegraphics[width=0.95\textwidth]{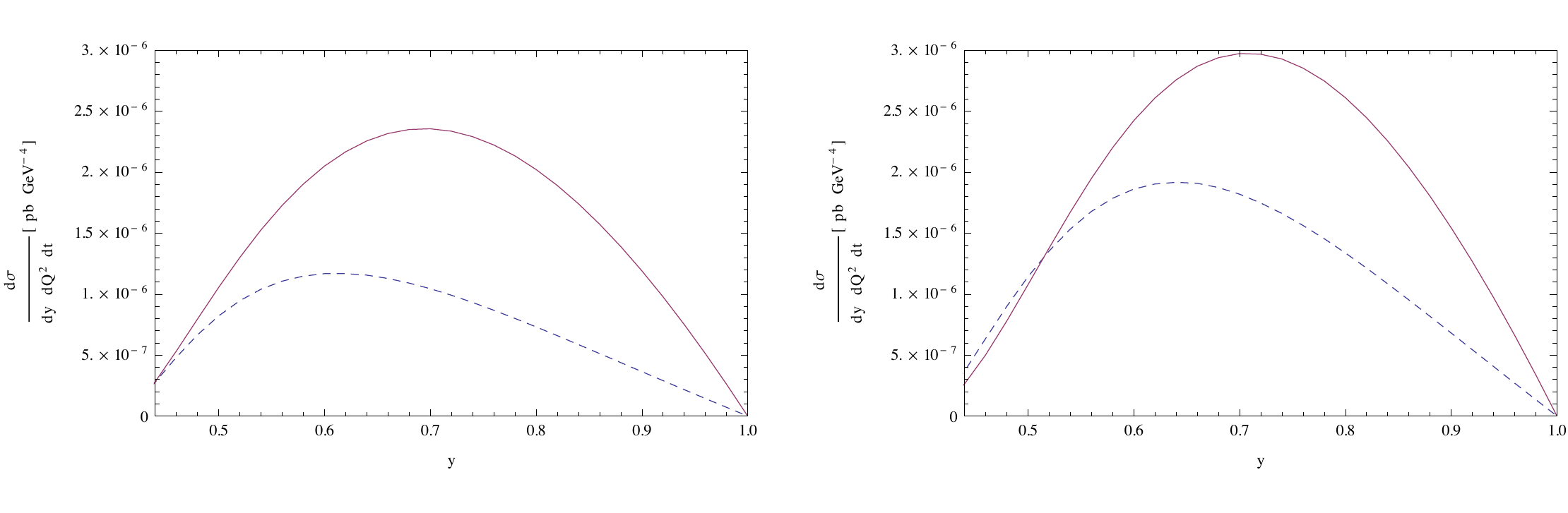}
\caption{The $y$ dependence  of the longitudinal contribution to the cross section $\frac{d\sigma(\nu N \to l^- N D^+)}{dy\, dQ^2\, dt}$ (in pb GeV$^{-4}$) for $Q^2 = 1$ GeV$^2$, $\Delta_T = 0$  and $s=20$ GeV$^2$ for a  proton (left panel) and neutron (right panel) target : total (quark and gluon, solid curve) and  quark only (dashed curve) contributions.}
   \label{neutrino_cs}
\end{figure}

Although hadronic physics is not today the main purpose of neutrino experiments which focus mostly on the study of neutrino oscillations, let us stress that data collected at the near detectors contain very valuable information which should be analyzed with care. For instance, it has been noticed \cite{Pire:2015iza} that the elusive chiral-odd quark transversity GPDs contribute to the exclusive $D$  meson production, thanks to the charmed mass effect in the quark propagators. Moreover, the gluon transversity GPDs contribute to the vector meson $D^*$ production when this charmed meson is transversely polarized \cite{Pire:2017yge}.  A complete study \cite{Pire:2017lfj} showed that the cross sections were small but yet sufficient for a dedicated study to be performed with an hopefully successful output, at least for the longitudinal contributions related to vector and axial GPDs. Without entering into details, we illustrate our results by showing in Fig. \ref{neutrino_cs} the differential cross-section $\frac{d\sigma(\nu N \to l^- N D^+)}{dy\, dQ^2\, dt}$. Note that the gluonic contribution is enhanced by a factor $\frac{V_{cs}^2}{V_{cd}^2}$ if one considers $D_s^+$ rather than $D^+$ production, with $V_{cq}$ the relevant CKM matrix elements.

\section{Prospects for EIC}
\label{sec:EIC}
Exclusive charged current reactions in high luminosity electron accelerators have been  studied with optimistic results \cite{Siddikov:2017nku, Siddikov:2019ahb}. We enlarged \cite{Pire:2021dad} this study by considering the production of charmed meson. Since the scattering amplitude is proportional to the relevant CKM matrix element, the dominant production process   involves a $D^-_s(1968)$ or  $D^{*-}_s(2112)$ charmed and strange meson. The $D^{*-}_s(2112)$ decays mostly in a $D^-_s(1968) \gamma$ pair.
Neglecting the strange quark content of the nucleon, the exclusive production of these mesons occur in the kinematical domain where collinear factorization  allows a description of the scattering amplitude as a convolution of gluon GPDs and the $D_s^-$ or $D_s^{*-}$ meson distribution amplitude (DA), 

\begin{figure}[h]
    \centering
\includegraphics[width=1\textwidth]{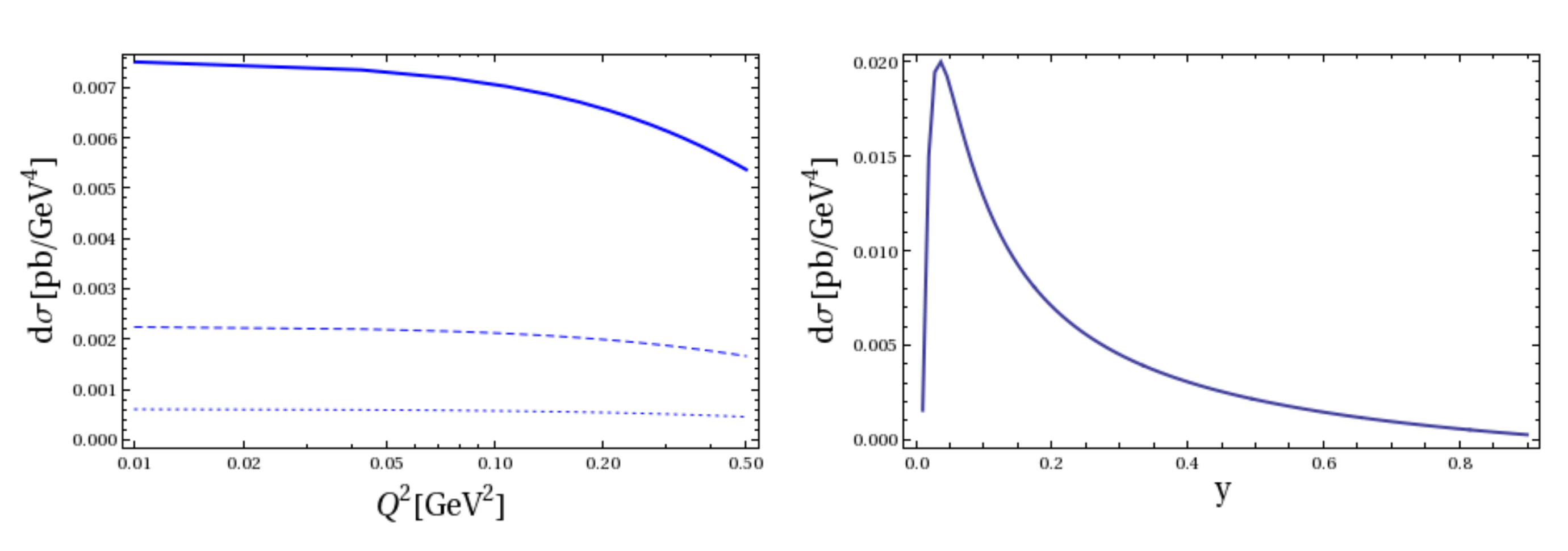}
    \caption{Left panel : The $Q^2$ dependence  of  the cross section $\frac{d\sigma(e^- N \to \nu N D_s^-)}{dy\, dQ^2\, dt}$ (in pb GeV$^{-4}$) for  $\Delta_T = 0$  and $s=820$ GeV$^2$ and $y=.2$ (solid curve) , $y=.5$ (dashed curve) and $y=.8$  (dotted curve). Right  panel : The $y$ dependence of  the cross section $\frac{d\sigma(e^- N \to \nu N D_s^-)}{dy\, dQ^2\, dt}$ (in pb GeV$^{-4}$) for $Q^2=0.1$ GeV$^2$, $\Delta_T = 0$  and $s=820$ GeV$^2$.}
    \label{figEIC0}
\end{figure}
\begin{figure}[h]
    \centering
\includegraphics[width=1\textwidth]{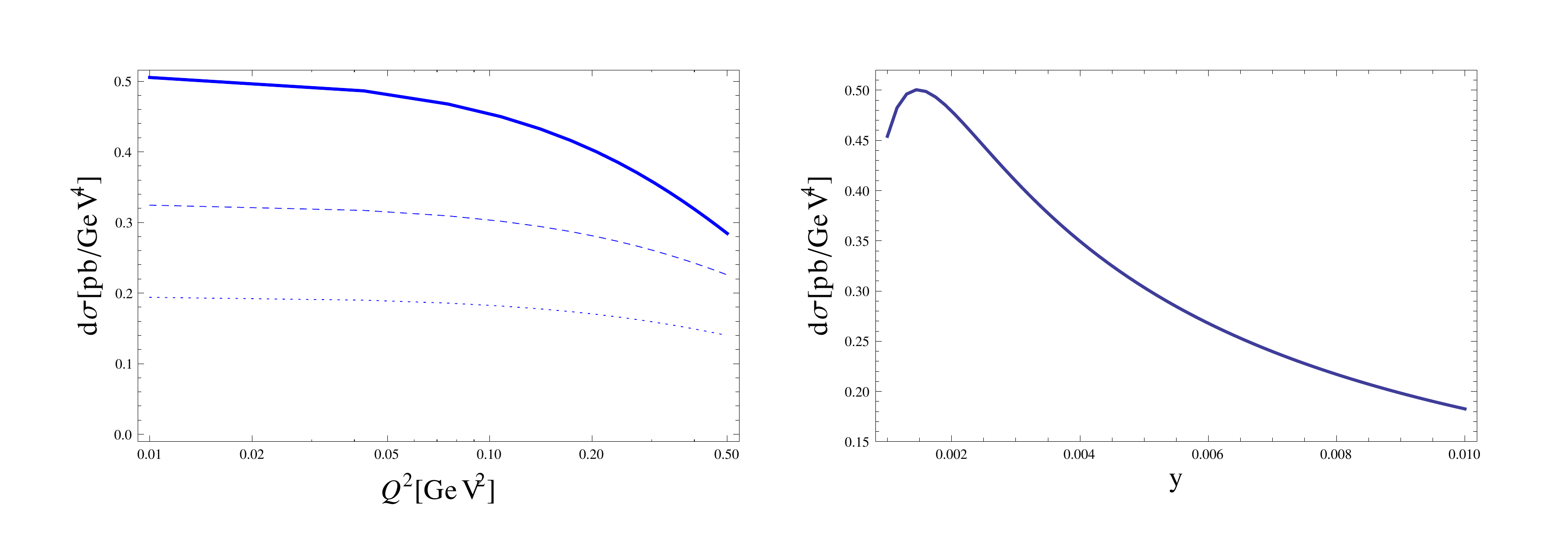}
    \caption{Left panel : The $Q^2$ dependence  of  the cross section $\frac{d\sigma(e^- N \to \nu N D_s^-)}{dy\, dQ^2\, dt}$ (in pb GeV$^{-4}$) for  $\Delta_T = 0$  and $s=20000$ GeV$^2$ and $y=10^{-3}$ (solid curve) , $y=5\cdot 10^{-3}$ (dashed curve) and $y=10^{-2}$  (dotted curve). Right  panel : The $y$ dependence of  the cross section $\frac{d\sigma(e^- N \to \nu N D_s^-)}{dy\, dQ^2\, dt}$ (in pb GeV$^{-4}$) for $Q^2=0.1$ GeV$^2$, $\Delta_T = 0$  and $s=20000$ GeV$^2$.}
    \label{figEIC1}
\end{figure}

\begin{figure}
    \centering
 \includegraphics[width=1\textwidth]{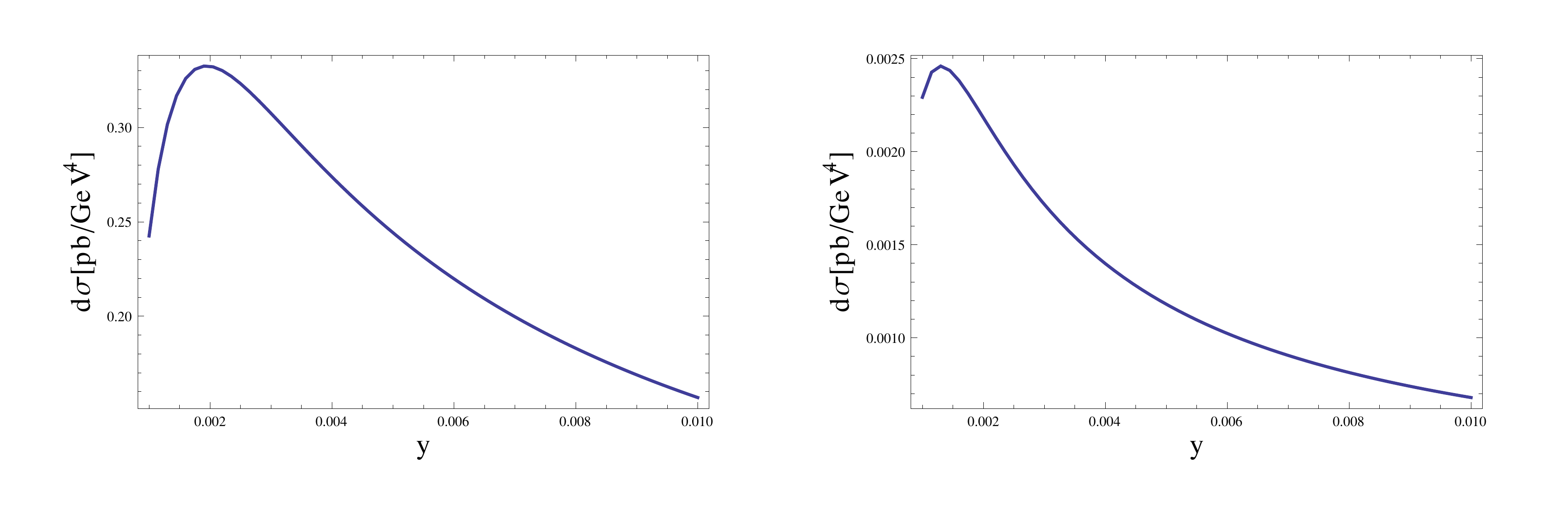}
    \caption{Left panel : The $y$ dependence  of  the longitudinal cross section $\frac{d\sigma(e^- N \to \nu N D_s^{-*})}{dy\, dQ^2\, dt}$ (in pb GeV$^{-4}$) for  $\Delta_T = 0$  and $s=20000$ GeV$^2$ and $Q^2=.1$ GeV$^2$ . Right  panel : idem for the production  cross-section of a transversely polarized $D_s^{-*}$.}
    \label{figEIC2}
\end{figure}
Our results, shown on Fig. \ref{figEIC0} and\ref{figEIC1} for the pseudoscalar $D_s$ meson and on Fig. \ref{figEIC2} for the two polarization modes of the vector $D^*_s$ meson are quite encouraging. Note that the kinematics of the solid curve on the left pannel of Fig. \ref{figEIC0} (resp.Fig. \ref{figEIC1} corresponds to $\xi = 0.012$ (resp. $\xi = 0.1$). The cross-section for the production of the transverse polarization mode of the $D^*_s$ is suppressed. As it may have been anticipated the $Q^2$ dependence is quite modest at small $Q^2 << M_D^2$. The $y$-dependence is quite strong resulting in the dominance of the moderate skewness region. The dependence of our results with respect to the choice of the gluon GPDs and heavy meson DAs is discussed in \cite{Pire:2021dad}.

\section{Conclusion}
The hadronic physics opportunities opened by  high luminosity and medium energy neutrino facilities have definitely be underexploited up to now. We demonstrated that the study of exclusive channels like $D$ or $D^*$ charmed meson production are very complementary to the hadronic program of electron facilities such as JLab, by giving access to elusive quantities like the quark or gluon transversity GPDS. We believe that planned high energy neutrino experiments such  as Minerve and Minos$+$ 
\cite{Ayres:2004js,Aliaga:2013uqz} which have their scientific program oriented toward the understanding of neutrino oscillations  will collect more statistics and will thus allow some nice progress in the realm of hadronic physics.

On the other hand, we have shown that the production cross-sections for charged current exclusive $D_s$ charmed strange mesons, although small, are in the reach of future high luminosity electron-ion colliders making them another potential source of information for future programs aiming at the extraction of GPDs. The rate for the longitudinally polarized $D_s^*$ vector meson is of the same order of magnitude as the one for the pseudoscalar $D_s$ meson. Both are in fact of the same order of magnitudes as the rates for light mesons at a $Q^2$ value of the order a few GeV$^2$ \cite{Siddikov:2019ahb}. A full feasibility study is much needed, since detecting a $D_s$ meson is not a trivial goal in a charged current event where the momentum of the outgoing neutrino cannot be measured.

\section*{Acknowledgements}
The work of J.W. is supported by the grant 2017/26/M/ST2/01074 of the National Science Center in Poland, whereas the work of L. S. is supported by the grant 2019/33/B/ST2/02588 of the National Science Center in Poland. This project is also co-financed by the Polish-French collaboration agreements Polonium, by the Polish National Agency for Academic Exchange and COPIN-IN2P3 and by the European Union’s Horizon 2020 research and innovation programme under grant agreement No 824093.




\nolinenumbers

\end{document}